\documentstyle[12pt]{article}
\begin{document}
\begin{titlepage}

\title{Asymptotic Limits and Structure of the Pion Form Factor}

\author{John F. Donoghue and  Euy Soo Na\\ [5mm]
Department of Physics and Astronomy\\
University of Massachusetts
Amherst, MA ~01003 ~U.S.A.}

\date{ }
\maketitle

\begin{abstract}

We use dispersive techniques to address the behavior of the pion 
form factor as $Q^2 \to \infty$ and $Q^2 \to 0$.
We perform the matching with the constraints of perturbative QCD and 
chiral perturbation theory in the high energy and low energy limits,
leading to four sum rules.
We present a version of the dispersive input which is consistent with
the data and with all theoretical constraints.
The results indicate that the asymptotic perturbative QCD limit is
approached relatively slowly, and give a model independent
determination of low energy chiral parameters.

\end{abstract}
{\vfill UMHEP-435; hep-ph/9611418}
\end{titlepage}

We are fortunate to have the rigorous techniques of perturbative
QCD[1] and chiral perturbation theory[2] describing the high and low
energy domains of the strong interactions respectively.
The only comparatively rigorous techniques that apply to the
intermediate energy region are lattice simulations[3] and dispersion
relations[4]. 
Dispersive techniques are increasingly being combined with the other
theoretical methods in order to provide as much control as possible
throughout all energy regions. The simpliest cases are the two point
functions of vector and axial vector currents[5],which are associated
with the Weinberg sum rules[6] and other related sum rules.
The next simpliest example is the three point function of the pion
electromagnetic form factor. The purpose of this paper is to discuss
the dispersive treatment of the pion form factor. We apply chiral
constraints at low energy and incorporate the behavior of perturbative
QCD at the highest energies. This leads to four sum rules, two of
which are reasonably obvious and two which are new. In addition the
dispersive treatment allows us to address the question of how fast the
form factor approaches the asymptotic QCD behavior[7].

The form factor is defined by
\begin{equation}
\langle \pi^+ \mid J^{em}_{\mu} \mid \pi^+ \rangle = f_{\pi}(q^2)
(p+p')_{\mu}
\end{equation}                                  
\noindent with $q_{\mu} = (p'-p)_{\mu}$ 
In its twice subtracted form, the dispersion relation for the pion
form factor reads

\begin{equation}
f_{\pi} (q^2) = 1 + Kq^2 + {q^4 \over \pi} \int^{\infty}_{4m^2_{\pi}} {ds
\over s^2} {Im f_{\pi} (s) \over s - q^2 - i
\epsilon}
\end{equation}                                   

\noindent Our results are independent of the number of subtractions, but this
form is most useful in presenting our techniques. We have imposed the
normalization constant $f_{\pi}(0) = 1$, and the constant K is a
subtraction constant to be determined below. 

At the high energy end, perturbative QCD tells us that the asymptotic
behavior of the pion formfactor[7], 
with $Q^2 = -q^2$, is

\begin{eqnarray}
f_{\pi}(Q^2) = 16 \pi {{\alpha}_s (Q^2) F^2_{\pi} \over Q^2} 
    = {64 {\pi}^2 \over 9} {F^2_{\pi} \over Q^2 ln {Q^2 \over {\Lambda}^2}} 
\end{eqnarray}                                            

\noindent with $F_{\pi} = 93 MeV$. \\

\noindent The fact that this decreases faster than $1/Q^2$ implies three
sum rules when combined with Eq 2. 
The fact that there is no term proportional to $Q^2$ as $Q^2 \to
\infty$ implies a sum rule for the subtraction constant

\begin{equation}
K = {1 \over \pi} \int^{\infty}_{4 m^2_{\pi}} {ds \over s^2} Im
f_{\pi} (s)
\end{equation}
                                      
\noindent Corresponding, there is no constant term as $Q^2 \to
\infty$, 
requires 
a sum rule which can be found by Taylor expanding the denominator at
large $Q^2$, yielding
              
\begin{equation}
1 = {1 \over \pi} \int^{\infty}_{4 m^2_{\pi}} {ds \over s} Im
f_{\pi} (s)
\end{equation}
                        
\noindent Finally, the lack of a $1/Q^2$ term in the asymptotic 
region implies that
 
\begin{equation}
0 = {1 \over \pi} \int^{\infty}_{4 m^2_{\pi}} ds Im
f_{\pi} (s)
\end{equation}                                    
 
\noindent These sum rules are contingent on the convergence of the integrals.
This is especially relevant for the last one, but we will see that the
integral is just barely convergent. 

At the low energy end, the pion form factor has been calculated to two
loops in chiral perturbation theory.
The result is

\begin{equation}
f_{\pi} (q^2) = 1 + {2 \bar {L_9} \over F^2_{\pi}} q^2 + c_V q^4 + O(q^6)
\end{equation}                                      

with
\begin{eqnarray}
\bar {L_9} &=& L^{(r)}_9(\mu) -{1 \over 192 {\pi}^2} \left( ln
{m^2_{\pi} \over {\mu}^2} + 1 \right) +{\bar {f_1} m^2_{\pi} \over 16
{\pi^2} F^2_{\pi}} \\ \nonumber
c_V &=& {1 \over 16 {\pi^2} F^2_{\pi}} \left \{ {1 \over 60 m^2_{\pi}} +
{\bar {f_2} \over 16 {\pi^2} F^2_{\pi}} \right \}
\end{eqnarray}

\noindent In this expression, the parameters $L^{(r)}_9(\mu)$ and
$c_V$, $\bar f_1$, $\bar f_2$ are 
renormalized parameters from the $E^4$ and $E^6$ chiral
Lagrangians, respectively.  $L^{(r)}_9(\mu)$ can in principle be in
other reactions, although it is most common to extract it from this
form factor. One give a dispersion sum rule for
$L^{(r)}_9(\mu)$ by expanding the chiral result around $q^2 =0$ to
find that $L^{(r)}_9(\mu)$ is related to the subtraction constant K
defined above. The precise relation is

\begin{eqnarray}
K = {2 \bar {L_9} \over F^2_{\pi}} = {2 L^{(r)}_9(\mu) \over F^2_{\pi}}
-{1 \over 96 {\pi}^2 F^2_{\pi}} \left( ln {m^2_{\pi} \over {\mu}^2} + 
1 \right) 
\end{eqnarray}                                        

\noindent Here and in what follows we drop reference to the chiral
constant $\bar f_1$ since its effect is so small due to the factor of
$m_\pi^2$ multiplying it in Eq. 8. Note that relation for ${\bar
{L_9}}$ is independent of the arbitrary scale $\mu$,
as the dependence of $L^{(r)}_9(\mu)$ on $\mu$ is compensated by the
explicit $\mu$ behavior displayed above. This exercise can be repeated
to find the term at order $Q^4$, both in the dispersion relation and
in the chiral expansion. The result is

\begin{eqnarray}
c_V  = {1 \over 16 {\pi}^2 F^2_{\pi}} \left\{ {1 \over 60 m^2_{\pi}} +
{\bar {f_2} \over 16 {\pi}^2 F^2_{\pi}} \right\} = 
{1 \over \pi} \int^{\infty}_{4 m^2_{\pi}} {ds \over s^3} Im
f_{\pi} (s)
\end{eqnarray}
                                         
For completeness, let us briefly describe how these sum rules would be
derived using an unsubtracted dispersion relation,

\begin{equation}
f_{\pi} (q^2) = {1 \over \pi} \int_{0}^{\infty} ds {Im f(s) \over s - q^2 - i
\epsilon}
\end{equation}
                                        
\noindent In this case, the sum rules of Eq. 5, 9, 10 all follows from
Taylor
expanding around $q^2 =0$, while Eq. 6 follows from the $q^2 \to
\infty$ limit. Dispersion relations connect the high and low energy
limits by providing constraints on  the whole analytic functions. It is
interesting that a given sum rule may follow from constraints on the 
high energy end in one
derivation yet emerge in the low limit in another approach. 

We now turn to the construction of a representation of $Im f_\pi(s)$
which is consistent with both the data and with theoretical
constraints. The easiest step is at low energy, where chiral symmetry
requires the structure[8]

\begin{equation}
Im f_{\pi} (s) = {s(1- {4 m^2_{\pi}\over s})^{3/2} \over 96 {\pi}^2
F_\pi^2}  {\theta} (s - 4
m^2_{\pi}) + O(s^2)
\end{equation}                                         

\noindent In the intermediate energy region, we have data on both the real and
imaginary parts of $f_\pi(s)$. There is nothing surprising here, the
physics is just that of the rho resonance. We take  $Im f_\pi(s)$ from
a fit to the data in Ref[9]. Matching with the low energy limit is
simple, as the resulting function is easily adjusted to approach Eq. 12
as $s \to 0$. 

For the high energy end, we need to choose an asymptotic form for 
$Imf_\pi(s)$ which yields Eq. 3 when inserted in a dispersion
relation. To see that this is the appropriate procedure, we consider
dividing the dispersive integral into two pieces, with the transition
part $\bar s$ being large enough that above $s = \bar s$ we are in
asymptotic high energy behavior for $Im f_\pi(s)$ 
              
\begin{equation}
f_{\pi} (q^2) = {1 \over \pi} \int_{0}^{\bar s} ds {Im f(s) \over s
+ Q^2} + {1 \over \pi} \int_{\bar s}^{\infty} ds {Im f(s) \over s
+ Q^2}
\end{equation}                             

\noindent In the first integral the integrand is finite and the range is finite
so that the result is analytic in $1/Q^2$ around $Q^2 \to \infty$.
As a consequence, the logarithm in the QCD form for the asymptotic
limit cannot be reproduced from the first integral, and must come from
the $s \to \infty$ behavior of $Im f_\pi(s)$ in the second integral.
The form of the imaginary part which guarantees the proper asymptotic
limit is

\begin{equation}
Im f_{\pi} (s) = - {64{\pi}^3 \over 9} {1 \over Q^2 ln^2 {Q^2 \over 
{\Lambda}^2}}
\end{equation}
                 
\noindent This form also allows the high energy sum rule Eq. 6 to converge.
A final step is the matching between the intermediate and high energy
forms of $Im f_\pi(s)$. In order to help with this we impose the high
energy sum rule Eq. 6 and the normalization integral Eq. 5.
For these to be satisfied, the negative values of $Im f_\pi(s)$
obtained from the asymptotic form at large $s$ must extend to fairly low
energies in order to be able to cancel the known positive contribution
effects of the rho. This is a powerful constraint. There is certainly
some ambiguity in the precise form in the matching region, but we have
found a relatively simple solution. This is depicted in Fig. 1,
showing a smooth matching slightly above 1 $GeV$. 

If we use this form for $Im f_\pi(s)$ in a dispersion relation, we
clearly have no predictive power in the intermediate energy region
where our method is data-driven. The predictions come from the
approaches to the asymptotic regions. Within a dispersive framework
the transitions to the low energy and high energy limits are both
determined largely by the numerically important intermediate energy
region. Our results are presented graphically in Fig. 2. On the low
energy side the structure of the real part of the form factor is
governed by the low energy constraints of Eq. 9,10. These are
predicted by the dispersion relations to have the form

\begin{eqnarray}
L_9^{(r)} (\mu) &=& {1 \over 192 \pi^2} \left( ln {m^2_{\pi} \over \mu^2} + 1
\right) + {F^2_{\pi} \over 2\pi} \int^{\infty}_{4 m^2_{\pi}} {ds \over s^2} Im
f_{\pi} (s) \\ \nonumber
  &=& 0.0074  \\ \nonumber
c_V &=& 4.1 GeV^{-4} \\ \nonumber
\bar {f_2} &=& 6.6
\end{eqnarray}   
                                          
\noindent using $\mu = m_\eta$. 
The result for $L_9^{(r)}$ agrees with the standard result,
derived from the
real part of the form factor. This is just a consistency condition
for the dispersion relation. Of greater conceptional interest is the
way that the dispersion method embodies the underlying physics of
vector meson dominance(VMD), and the way that it resolves the issue of
the scale dependence of the chiral coefficients in VMD[10]. 
Vector dominance is motivated by a narrow width approximation to the
dispersion integral

\begin{equation}
Im f_{\pi} (s) = {\pi m_\rho^2} \delta (s - m^2_{\rho})
\end{equation}
Ecker, Pich and de Rafael argued that VMD determines the chiral
coefficients at the scale ${\mu}^2 = m^2_{\rho}$. The dispersive
approach provides a different answer --- VMD determines not simply the
chiral coefficient $L_9^{(r)} (\mu)$ but rather the scale independent
combination of the coefficient plus a specific combination of chiral
logs, i.e. ${\bar {L_9}}$ in Eq. 9.

On the high energy side, we see from Fig. 2 that the asymptotic QCD limit is
approached rather slowly.  In a dispersive framework this is due to
the large contributions of the soft physics region, most notably the
rho resonance, which continues to be more important in the dispersion
integral than the somewhat small perturbative contribution. This
result is consistent with quark model calculations[11], but is far
less model dependent.

The techniques of dispersion relations provide a partial bridge
between the low energy techniques of chiral perturbation theory and
the high energy techniques of QCD. The simplest exploration of these
methods involve two point functions. The present work involves a three
point function and hence is a step towards the consideration of yet
more difficult matrix elements such as the nonleptonic amplitudes
responsible for electromagnetic mass difference[12] or weak decays.

{\bf References}.\\
1) For a recent survey, see {\bf QCD - 20 Years Later}, ed by P.M.
Zerwas and H.A. Kastrup (World Scientific, Singapore, 1993). \\
2) e.g. see J. F. Donoghue, E. Golowich and B. R. Holstein, {\it Dynamics of
the Standard Model}, (Cambridge University Press, Cambridge, 1992).\\
3) A recent review is in Proc. XIV Intl. Conf. on Lattice Physics,
(Elsevier Press, Amsterdam, 1996). \\
4)J. F. Donoghue, in {\it Chiral
Dynamics of Hadrons and Nuclei}, ed by D.P. Min and M. Rho (Seoul
National University Press, Seoul, 1995), p87 (hep-ph/9506205) and
hep-ph/9607351.\\ 
R. Kronig, J. Op. Soc. Am. {\bf 12}, 547 (1926); H. A. Kramers,
Atti Cong. Int. Fisici Como (1927).\\
G. Barton, {\it Dispersion Techniques in Field Theory}, (Benjamin, NY,
1965).\\
5)J. Gasser and H. Leutwyler, Nucl. Phys. {\bf B250}, 465 (1985).\\  
J. F. Donoghue and E. Golowich, Phys. Rev. {\bf D49}, 1513
(1994).\\
6) S. Weinberg, Phys. Rev. Lett. {\bf 17}, 616 (1966).\\ 
T. Das, V. Mathur, and S. Okubo, Phys. Rev. Lett. {\bf 19}, 859
(1967).\\
T. Das, G. S. Guralnik, V. S. Mathur, F. E. Low, and J. E. Young,
Phys. Rev. Lett. {\bf 18}, 759 (1967).\\
J. F. Donoghue and E. Golowich, Phys. Lett. {\bf 315}, 406
(1993).\\
J. F. Donoghue and E. Golowich, Ref 5.\\
7)S. J. Brodsky and G. P. Lepage, Phys. Rev.{\bf D22}, 2150 (1980).\\
A. Duncan and A. Mueller Phys. Rev. {\bf D21}, 1636 (1980).\\
8)J. Gasser and H. Leutwyler, Ref 5.\\
J. Gasser and U. G. Meissner, Nucl Phys. B357, 90 (1991).\\
9) A. Bericha, G. Lopez Castro, and J. Pestieau, Phys. Rev. {\bf D53},
4089 (1996). \\
10) J. F. Donoghue, C. Ramirez and G. Valencia, Phys. Rev. {\bf D39},
1947 (1988), \\
G. Ecker, J. Gasser, A. Pich and E. de Rafael, Nucl. Phys. {\bf B321},
311 (1989).\\
11) N. Isgur and C. Llewellyn-Smith, Phys. Rev. Lett. {\bf 52}, 1080
(1988), Nucl. Phys. {\bf B317}, 526 (1989).\\
L. S. Kisslinger and S. W. Wang, hep-ph/9403261 .\\
12) J. F. Donoghue and A. Perez, hep-ph/9611331, \\
J. Bijnens and J. Prades, hep-ph/9610360. \\ 

\vspace{1.5cm}

\noindent {\bf Figure Captions}\\
Fig. 1. Fit to the imaginary part of the pion form factor satisfying
the consistency constraints described in the text. \\
Fig. 2. The real part of the pion form factor at large $Q^2$. The
dashed line indicates the asymptotic prediction of perturbative QCD
with $\Lambda = 0.3 GeV$. \\

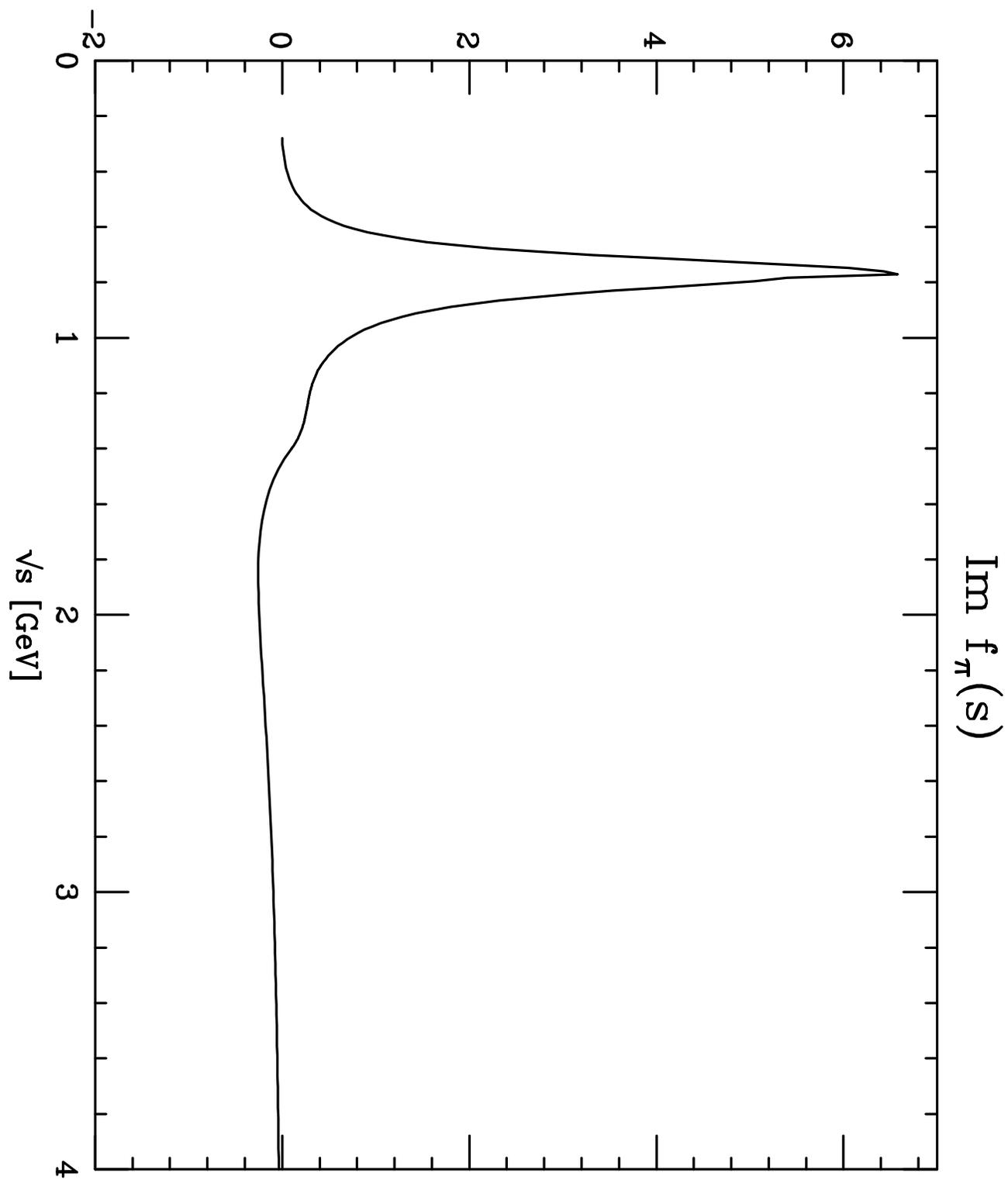
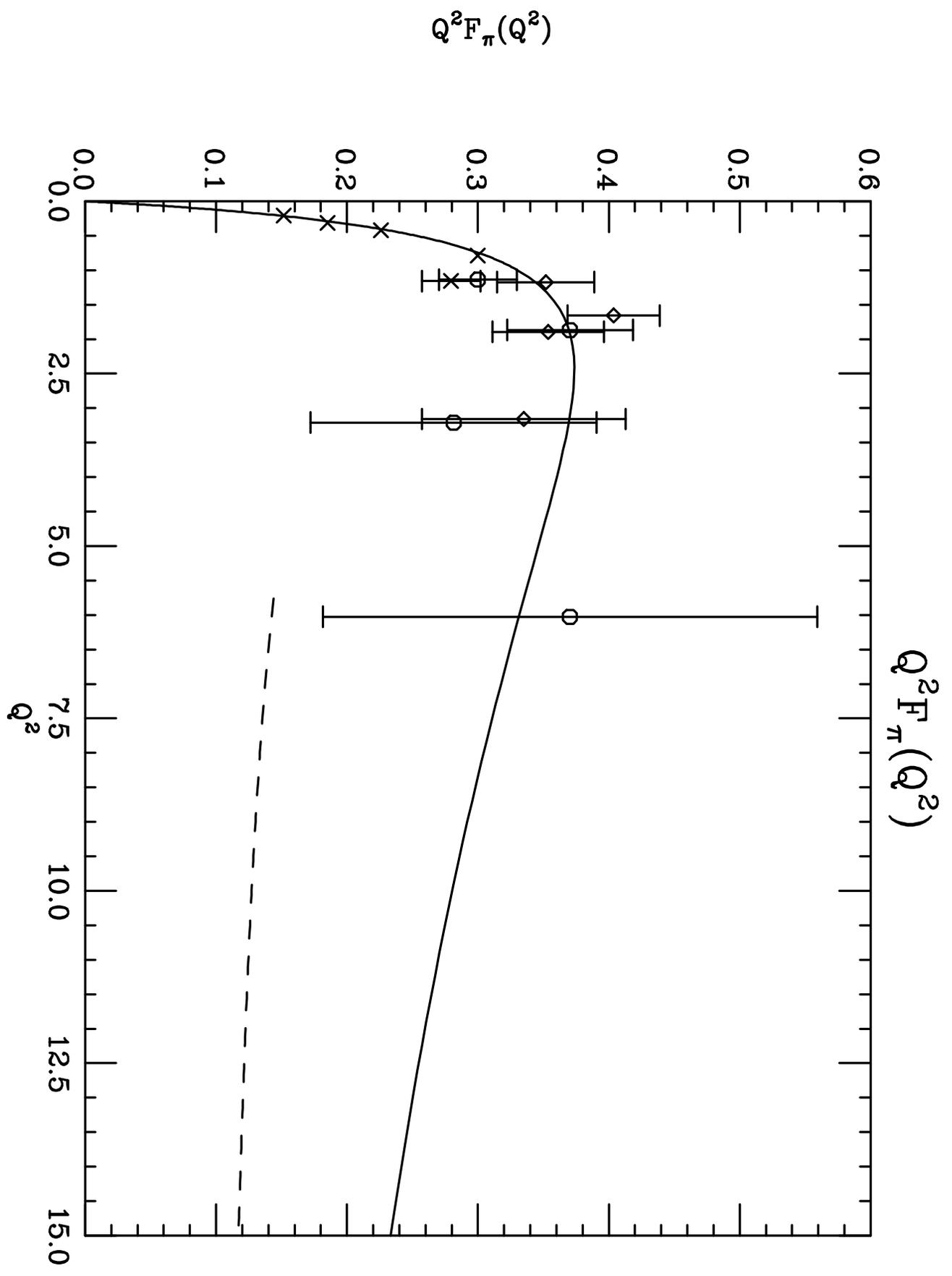
\end{document}